\documentclass[seceq]{ptptex}

\usepackage{graphicx}

\title{Deconvolution of window effect in galaxy power spectrum analysis}

\author{
\textsc{Takahiro Sato}${}^1$, 
\textsc{Gert H\"utsi}${}^{2}$,
\textsc{Kazuhiro Yamamoto}${}^1$}

\inst{
$^1$Department of Physical Sciences, Hiroshima University, 
Higashi-hiroshima, 739-8526, Japan
\\
$^{2}$Tartu Observatory, EE-61602 T\~oravere, Estonia
}

\date{\today}

\abst{
We develop a new method for deconvolving the smearing effect of 
the survey window 
in the analysis of the galaxy multipole power spectra from a redshift survey.
This method is based on the deconvolution theorem, and is 
compatible with the use of the fast Fourier transform. 
It is possible to measure the multipole power spectra 
deconvolved from the window effect efficiently.
Applying this method to the luminous red galaxy sample of the 
Sloan Digital Sky Survey data release 7 as well as mock catalogues, 
we demonstrate how the method works properly.
Using this deconvolution technique, the amplitude of the multipole 
power spectrum is corrected. Besides, the covariance matrices of 
the deconvolved power spectra get quite close to the diagonal form. 
This is also advantageous in the study of the BAO signature. 
}

\begin{document}
\maketitle

\def\bfk{{\bf k}}
\def\bfs{{\bf s}}

\section{Introduction}
The power spectrum is one of the most fundamental tools to characterize 
the large-scale clustering of galaxies (e.g., \citen{Dodelson}).
The role of the power spectrum analysis in cosmology is 
becoming more and more important \cite{Tegmark06}. 
Especially, the power spectrum of galaxies plays a key role for 
extracting the baryon acoustic oscillations (BAO) \cite{Hutsia,Hutsib,Percival}, 
which is essential for the dark energy research. 
The quadruple power spectrum is also useful for measuring the 
redshift-space distortion, which plays a vital role of 
testing gravity on the scales of cosmology \cite{Linder,Guzzo,Yamamoto08}. 

The method developed by Feldman, Kaiser and Peacock (hereafter FKP, 
\citen{FKP94}) is often adopted to measure the power spectrum \cite{Percival}.
It is known that the power spectrum measured using the FKP method,
due to finiteness of the survey volume, is convolved with a window function.
This convolution is expressed schematically as,
\begin{eqnarray}
P^{\rm conv}(\bfk)={1\over (2\pi)^3}\int d^3k' P(\bfk')W(\bfk-\bfk'),
\label{convone}
\end{eqnarray}
where $\bfk$ is the wavenumber vector, $W(\bfk-\bfk')/(2\pi)^3$ 
describes 
the survey window, and we refer to $P^{\rm conv}(\bfk)$ as the `convolved' power 
spectrum. In the limit of an infinitely large survey volume, 
$W(\bfk-\bfk')/(2\pi)^3$
approaches the three dimensional delta function, hence
$P^{\rm conv}(\bfk)$ becomes the same as the true power spectrum 
$P(\bfk)$. 

In a realistic case, however, a survey volume is finite. 
The window convolution modifies a measured power spectrum compared 
with the true power spectrum.
It has the following two aspects: 
(i) change of the amplitude of the power spectrum, 
(ii) mixing of modes with different wavenumbers. These two 
effects are influential in the power spectrum analysis for 
extracting the BAO and in testing gravity 
on the scales of cosmology.
When comparing the convolved power spectrum with theoretical predictions, 
we need to take the window effect into account 
following Eq.~(\ref{convone}), even though it is quite a 
time consuming process \cite{SatoW2}. 

If we could directly measure a deconvolved power spectrum from a galaxy data, 
it could be compared with 
theoretical models without including the window effect \cite{Fisher93}, 
thus speeding up the analysis significantly. 
In the present paper, for the first time, we develop a scheme to measure 
the deconvolved multipole power spectra. 
We apply it to the Sloan Digital Sky Survey (SDSS) 
luminous red galaxy (LRG) sample from the data release (DR) 7
as well as to $1000$ mock catalogues mimicking the LRG sample.
This paper is organised as follows: In section 2, we present a 
new method to deconvolve the window function from the FKP estimator, 
including a brief review for the FKP estimator for the power spectrum.
In section 3, we apply the method to the SDSS LRG DR7 as well as to
$1000$ mock catalogues,
and demonstrate how it works in the multipole power spectrum 
analysis, by comparing the convolved power spectrum and the 
deconvolved power spectrum.
Throughout this paper, we use units in which the velocity of 
light equals 1, and adopt the Hubble parameter $H_0=100h$ km/s/Mpc
with $h=0.7$.

\section{Formulation}
Here we provide basic formulas for the deconvolution of the window function.
Taking the Fourier transformation of Eq.~(\ref{convone}), we have
\begin{eqnarray}
\int d^3{k}
e^{-i\bfk\cdot\bfs} P^{\rm conv}(\bfk)
={1\over (2\pi)^3}
\left[\int d^3k'e^{-i\bfk'\cdot\bfs}P(\bfk')\right]
\left[\int d^3k e^{-i\bfk\cdot\bfs}W(\bfk)\right].
\end{eqnarray}
This can be done exactly as long as the k-space is infinitely large. 
Then, we have 
\begin{eqnarray}
\int d^3k'e^{-i\bfk'\cdot\bfs}P(\bfk')=(2\pi)^3{
\displaystyle{\int d^3{k'}
e^{-i\bfk'\cdot\bfs} P^{\rm conv}(\bfk')}\over
 \displaystyle{
\int d^3k'' e^{-i\bfk''\cdot\bfs}W(\bfk'')}},
\end{eqnarray}
whose inverse transformation yields
\begin{eqnarray}
P(\bfk)=\int d^3s e^{i\bfk\cdot\bfs} {
\displaystyle{\int d^3{k'}
e^{-i\bfk'\cdot\bfs} P^{\rm conv}(\bfk')}\over
 \displaystyle{
\int d^3k'' e^{-i\bfk''\cdot\bfs}W(\bfk'')}}.
\label{decp}
\end{eqnarray}

For a discrete density field, we need to subtract the shot-noise contribution. 
Following the FKP method, the estimator of the convolved power spectrum 
is 
\begin{eqnarray}
P^{\rm conv}(\bfk)={\bigl|\int d^3s \psi(\bfs)
[n_{\rm g}(\bfs)-\alpha n_{\rm rnd}(\bfs)] e^{i\bfk\cdot\bfs}\bigr|^2
\over
[\int d^3s \bar n^2(\bfs) \psi^2(\bfs)]}
-(1+\alpha)S_0,
\label{exp}
\end{eqnarray}
where $n_{\rm g}(\bfs)$ is the number density field of galaxies whose mean 
number density is $\bar n(\bfs)$, $n_{\rm rnd}(\bfs)$ is 
the density field of a random sample whose mean number density is 
$\alpha^{-1} \bar n(\bfs)$, $\psi(\bfs)$ is the weight factor, 
$\bfs$ denotes the three dimensional coordinate of the redshift-space,
and we defined 
\begin{eqnarray}
&&S_0=
{\int d^3s \bar n(\bfs) \psi^2(\bfs)\over\int d^3s \bar n^2(\bfs) \psi^2(\bfs)},
\label{exs}
\end{eqnarray}
which describes the shot-noise contribution.  
The random catalogue is a set of random points without any correlation,
which can be constructed through a random process by mimicking 
the selection function of the galaxy catalogue.
Note that $\alpha$ is 
the parameter of the random catalogue, for which we adopt 
the value of $0.01$, in the present paper.
Similarly, we may adopt the estimator for the window function 
\begin{eqnarray}
W(\bfk)={\big|\int d^3s \alpha n_{\rm rnd}(\bfs) \psi(\bfs) e^{i\bfs\cdot\bfk}\big|^2
\over \int d^3s \bar n^2(\bfs) \psi^2(\bfs)}
-\alpha S_0.
\label{exw}
\end{eqnarray}
Note that the denominator $\int d^3s \bar n^2(\bfs) \psi^2(\bfs)$ 
cancels out when we substitute Eqs.~(\ref{exp}), (\ref{exs}) and 
(\ref{exw}) into Eq.~(\ref{decp}). Practically, the computation
of $W(\bfk)$ must be done carefully, because it might become zero when 
we choose a box size for the fast Fourier transform too large. 
One can avoid this problem by properly choosing the box size 
in which sample galaxies are distributed.

The monopole power spectrum is computed from Eq.~(\ref{decp}),
\begin{eqnarray}
P_0(k)={1\over V_{k}} \int_{V_k} d^3k P(\bfk),
\end{eqnarray}
where $V_k$ is the volume of a shell in the Fourier space, whose 
mean radius is $k$. 
Within the small-angle approximation (distant observer approximation),
the estimator of the higher multipole power spectra might be taken as
\begin{eqnarray}
P_\ell(k)={1\over V_{k}} \int_{V_k} d^3k
P(\bfk) {\cal L}_\ell({\bf \hat e}\cdot{\bf \hat k}),
\label{Pell}
\end{eqnarray}
where $ \hat {\bf e}$ denotes the unit vector that points 
to line of sight direction, ${\cal L}_\ell(\mu)$ is the Legendre polynomial,
and $\hat{\bf k}=\bfk/|\bfk|$ is the unit wavenumber vector. 
Note that the line of sight direction is approximated by 
one direction $\hat{\bf  e}$ in this approach. 
We refer to the multipole power spectrum $P_{\ell}(k)$ with
Eq.~(\ref{decp}) substituted into the right hand side of (\ref{Pell}) 
as the `deconvolved' multipole power spectrum.
Note that our definition of the multipole spectrum 
$P_\ell(k)$ is different from the conventional one by the factor 
$2\ell+1$ \cite{CFW,Hamilton}.

In the ref.~\citen{YNKBN}, the FKP method is generalized to measure higher 
order multipole power spectra, which is useful in quantifying the 
redshift-space distortions (cf.~\citen{CFW,Hamilton}).
There are differences between the previous paper \cite{YNKBN} and 
the present one in the estimators for the higher multipole
spectra. In Table I, the 
differences of the two approaches are summarized. 
In the previous paper \cite{YNKBN}, the line of sight direction 
was defined for each pair of galaxies, while in the present paper
the line of sight direction is approximated by only one direction 
$ \hat{\bf e}$. In this sense, the approximation of the present
paper is limited to a set of galaxies within a narrow survey area. 
A similar approach was adopted in the references \citen{OutramI,OutramII}.
In this approach, because of the limitation of the small-angle approximation, 
a full sample of large survey area must be divided into subsamples 
of narrow survey areas. 
Because of the division of the full sample, the window effect could 
be very influential, but the advantage of this approach is that one
can use the fast Fourier transform.

\begin{table}[htbp]
 \begin{center}
  \begin{tabular}{l|ll}
   \hline
   ~ &  Present paper & Reference \citen{YNKBN} \\
   \hline\hline
   Line of sight direction & approximated by  &  
   computed for each \\
   ~~ & ~~one direction $\hat {\bf e}$ &  
   ~~pair of galaxies\\
   \hline
   Division of galaxy sample & necessary & not necessary  \\
   \hline
   Window effect & significant & not significant\\
   \hline
   Use of the fast Fourier transform & possible & impossible\\
   \hline
  \end{tabular}
  \caption{Comparison of the two approaches}
 \end{center}
\end{table}

\section{Application}

In this section we apply our method to the SDSS LRG sample of DR7.
Our LRG sample is restricted to the redshift range $z=0.16$ - $0.47$. 
In order to reduce the sidelobes of the survey window we 
remove some non-contiguous parts of the sample, which leads
us to 7150 deg${}^2$ sky coverage with the total number $N=100157$ LRGs.
The data reduction is the same as that described in the references 
\citen{Hutsia,Yamamoto10}.
In our power spectrum analysis we adopt the spatially flat lambda
cold dark matter model distance-redshift relation with $\Omega_m=0.28$, 
and set $\psi=1$.

We divide the full sample into subsamples, in which galaxies are 
distributed within narrow area. This is necessary to compute the 
quadrupole power spectrum in our method with the fast Fourier transform.
In the present paper, we divide the full sample 
into $18$ subsamples whose area and numbers of galaxies are almost equal,
$397$ square degrees, or into $32$ subsamples whose mean area is $223$ 
square degrees.

\subsection{Shape of the multipole power spectrum}

\begin{figure}[ht]
\begin{center}
\includegraphics[width=88mm, height=88mm]{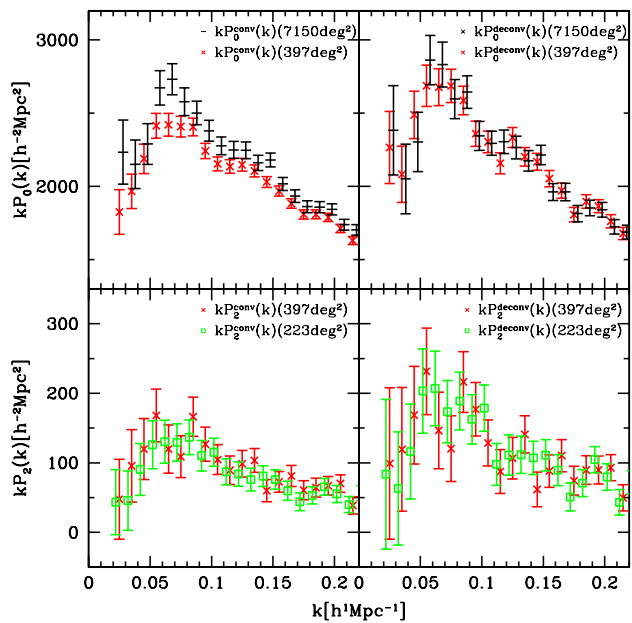}
\end{center}
\caption{
Comparison of the convolved $kP_\ell^{\rm conv}(k)$ (left panel) and 
deconvolved power spectra $kP_\ell^{}(k)$ (right panel).
The upper panels plot the monopole $\ell=0$,
while the lower ones the quadrupole $\ell=2$. 
For the monopole power spectrum, the two cases are presented: One is 
the case when the full sample is divided into $18$ subsamples and the other
is the case without division of the full sample. 
For the quadrupole spectrum, the two different cases of the division 
of the full sample are presented, in which the mean survey areas are
$397$ square degrees and $223$ square degrees, respectively.
The error-bars are obtained by computing the standard 
deviations of 1000 mock catalogues (A).
\label{fig:kpsdss}
}
\end{figure}

First, we focus on the shapes of the monopole and the quadrupole
power spectra. 
Figure \ref{fig:kpsdss} shows the {\it convolved} (left panels) and 
{\it deconvolved} (right panels) power spectra from the SDSS LRG sample.
The upper left panel plots the {\it convolved} monopole spectrum 
multiplied by the wavenumber, 
$kP^{\rm conv}_0(k)$.
Here the following two cases are shown: 
(i) the full sample divided into 18 subsamples
with each piece spanning almost $397$ square degrees,
(ii) the full sample without division into subsamples. 
Due to significantly different effect of the window convolution
in these two cases, the amplitude of the two spectra
are very different. 
The upper right panel is the same as the upper left panel, 
but is the {\it deconvolved} power spectrum $kP_0(k)$. 
Note that, in this panel, the amplitudes of the two 
deconvolved spectra are almost the same. The lower panels 
plot $kP_2(k)$. The left panel is the convolved spectrum, 
while the right panel is the deconvolved spectrum. 
For the quadrupole power spectrum, we present the two 
cases with different divisions: chunks with mean survey area
$397$ square degrees, and $223$ square degrees, respectively.
One can see the same features as those for $kP_0(k)$. 
The results do not depend on the division of the full sample. 
The error-bars are obtained by computing the standard 
deviations of 1000 mock catalogues (A), explained in the next.

\begin{figure}[ht]
\begin{center}
\includegraphics[width=88mm, height=88mm]{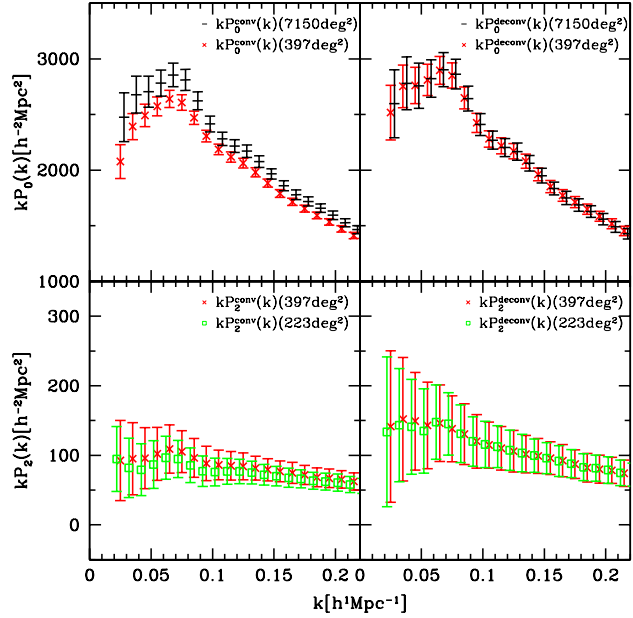}
\end{center}
\caption{
Same as figure \ref{fig:kpsdss} but using the average from $1000$ mock
catalogues (A).
The error-bars are the standard deviations.
\label{fig:kpsdsskpmock}
}
\end{figure}

Figure \ref{fig:kpsdsskpmock} is the same as figure 
\ref{fig:kpsdss}, but shows the average from 1000 mock 
catalogues (A), which are 
obtained by the following three steps, whose details are 
described in the reference \citen{Hutsia}.
First, we generate the density field using an optimized 2nd
order Lagrangian perturbation calculation. Second, we take
the Poisson sampling of the generated density field, adjusting
the clustering bias and the number density so as to agree
with the observed LRG sample. Third, we apply the radial 
and angular selection function, and extract the final 
catalogues. We call these mock catalogues (A).
Figure \ref{fig:kpsdsskpmock} clearly shows that the amplitudes 
of the two deconvolved spectra are almost the same.

\begin{figure}[t]
\begin{center}
\includegraphics[width=88mm, height=88mm]{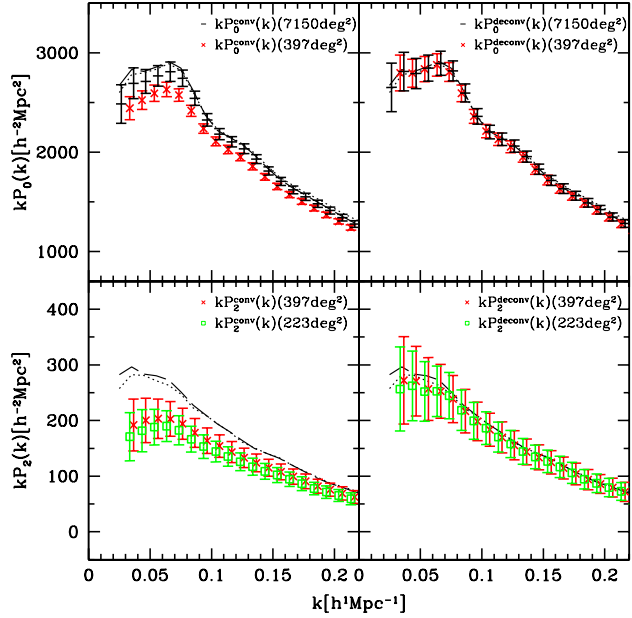}
\end{center}
\caption{
Same as figure \ref{fig:kpsdss} but using the average from the 
$1000$ catalogues (B). The solid curve and the dotted curve 
are the deconvolved spectrum and the convolved spectrum from 
the catalogues (C). 
\label{fig:kpsdsskpmockb}
}
\end{figure}

Figure \ref{fig:kpsdsskpmockb} demonstrates how the 
deconvolution recovers the original power spectrum properly. 
To this end, we buid other catalogues (B) and (C). 
The cosmological parameters are the same as those of 
the mock catalogues (A). 
To buid the catalogues (B), the first step is the same
as those for the mock catalogues (A), 
we generate the density field using an optimized 2nd
order Lagrangian perturbation calculation. 
Then, we take the Poisson sampling of the generated 
density field, adjusting the clustering bias and the 
number density. Here, we assumed the homogeneous mean 
number density with $\bar n=8\times 10^{-5} ~(h^{-1}{\rm Mpc})^{-3}$ 
in the square region $2560h^{-1}{\rm Mpc}\times2560h^{-1}{\rm Mpc}
\times1280h^{-1}{\rm Mpc}$.
The catalogues (B) are obtained by applying the same 
selection of the survey region as the SDSS LRG sample.
Hence, (B) has the same survey region as (A) but with the 
constant mean number density.

We also consider the catalogues (C), which are the same as 
(B) but without the selection of the survey region. Hence, 
each catalogue of (C) has the square region 
$2560h^{-1}{\rm Mpc}\times2560h^{-1}{\rm Mpc}\times1280h^{-1}{\rm Mpc}$. 
To obtain the redshift-space catalogues, however, 
for the catalogues (C), we assume an observer at an infinite 
distance away, in order to guarantee the validity of the 
distant observer approximation. Thus the catalogues (C) 
are ideal samples that assume the large volume and 
the validity of the distant observer approximation,
which could be expected to give us the original power spectrum. 

Figure \ref{fig:kpsdsskpmockb} plots the convolved spectra
(left panels) and the deconvolved spectra (right panels) 
from the catalogues (B). 
The upper panels are the monopole, while the lower 
panels the quadrupole spectra. 
In each panel, the convolved spectrum (dotted curve)
and the deconvolved spectrum (solid curve) from the ideal 
catalogues (C) are shown for comparison.
These two curves are almost same, hence the convolved spectrum 
and the deconvolved spectrum are almost same, because the 
catalogues (C) are ideal samples with the large volume. 
In the right panels of Fig.~\ref{fig:kpsdsskpmockb}, the good 
agreement of all the spectra indicates that our deconvolution 
recovers the original power spectrum properly. 
However, one can see small deviation in the quadrupole 
spectrum between the result from the catalogues (B) and (C), 
in the lower right panel. 
As the deviation is larger at small $k$, one possible reason  
of this small deviation is the limitation of the distant 
observer approximation for the catalogue (B).

\subsection{Covariance matrix}
In the following we discuss the effect of the deconvolution on the 
covariance matrix, which is defined by 
\begin{eqnarray}
{C}_{\ell\ell'}(k_i,k_j)
=\left<\Delta P_\ell(k_i)\Delta P_{\ell'}(k_j)\right>.
\label{covarinceFKP}
\end{eqnarray}
We obtain the covariance matrix $C_{\ell\ell'}$ using $1000$ mock catalogues. 
The correlation matrix, which describes the 
correlation of the errors between different
wavenumbers, is defined 
\begin{eqnarray}
r_{\ell\ell'}(k_i,k_j)={C_{\ell\ell'}(k_i,k_j)\over \sqrt {C_{\ell\ell}(k_i,k_i)
C_{\ell'\ell'}(k_j,k_j)}}.
\label{correlationmatrix}
\end{eqnarray}

\begin{figure}[ht]
\begin{center}
\begin{tabular}{ccc}
    \includegraphics[width=0.22\textwidth,angle=270]{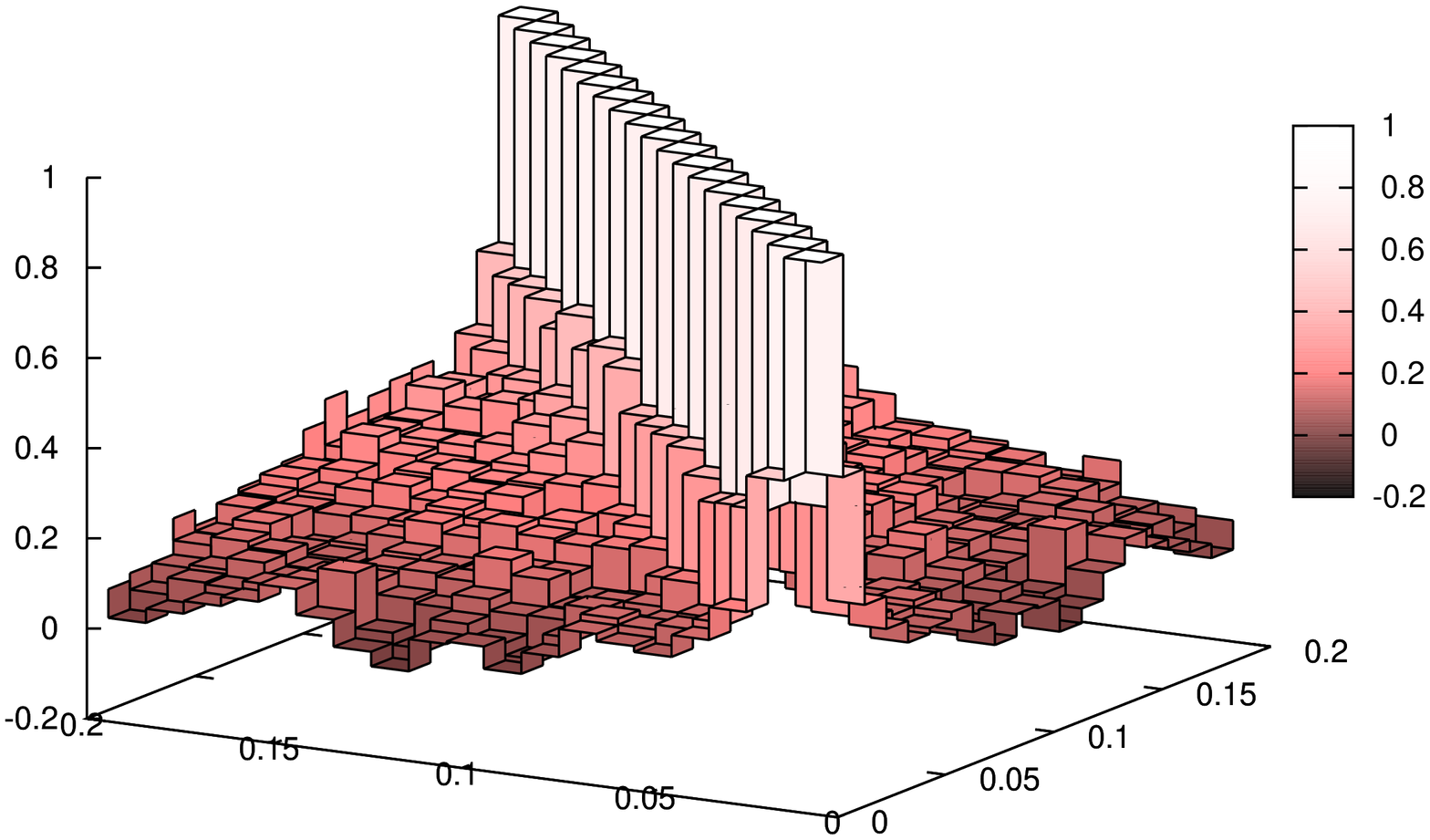}
&
    \includegraphics[width=0.22\textwidth,angle=270]{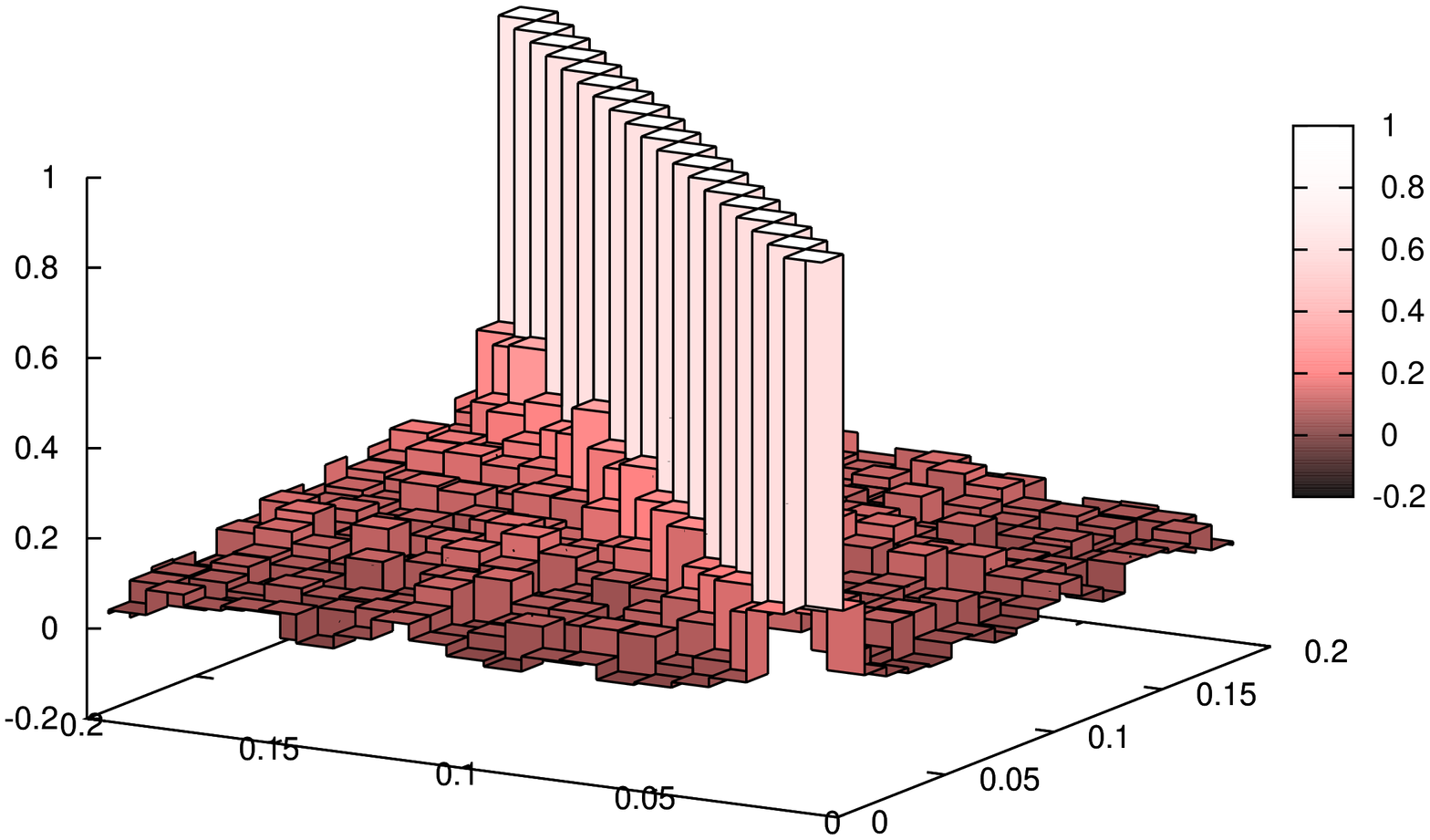}
&
    \includegraphics[width=0.22\textwidth,angle=270]{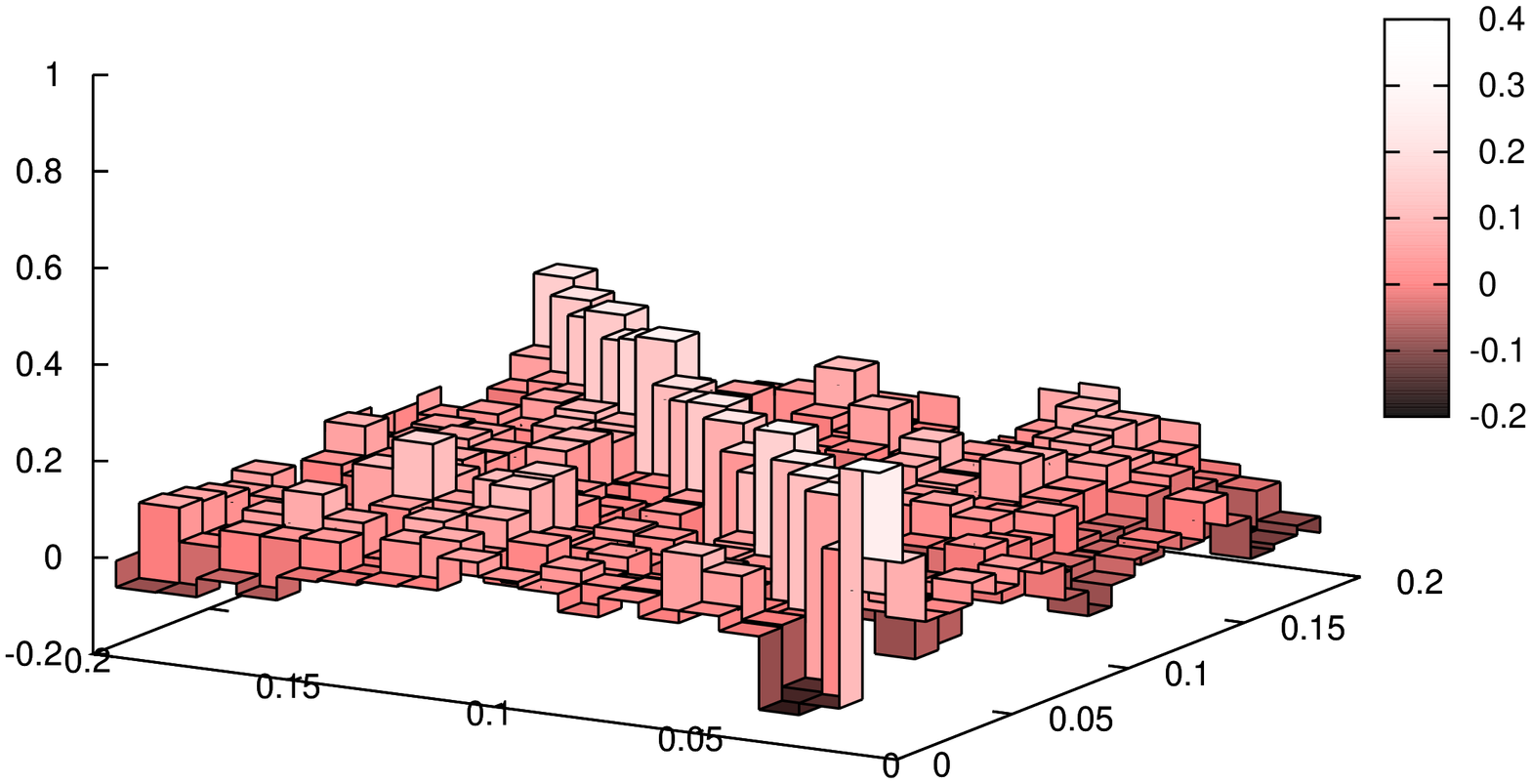}
\end{tabular}
\end{center}
\caption{
The correlation matrices defined by Eq.~(\ref{correlationmatrix}),
obtained from $1000$ mock catalogs:
$r_{00}(k,k')$ (left panel),
$r_{22}(k,k')$ (center panel)
and $r_{02}(k,k')$ (right panel),
respectively.
Here we used the {\it convolved} power spectra of 
the subsamples with the mean survey area $397$ deg${}^2$.
\label{fig:corrP0}
}
\begin{center}
\begin{tabular}{ccc}
    \includegraphics[width=0.22\textwidth,angle=270]{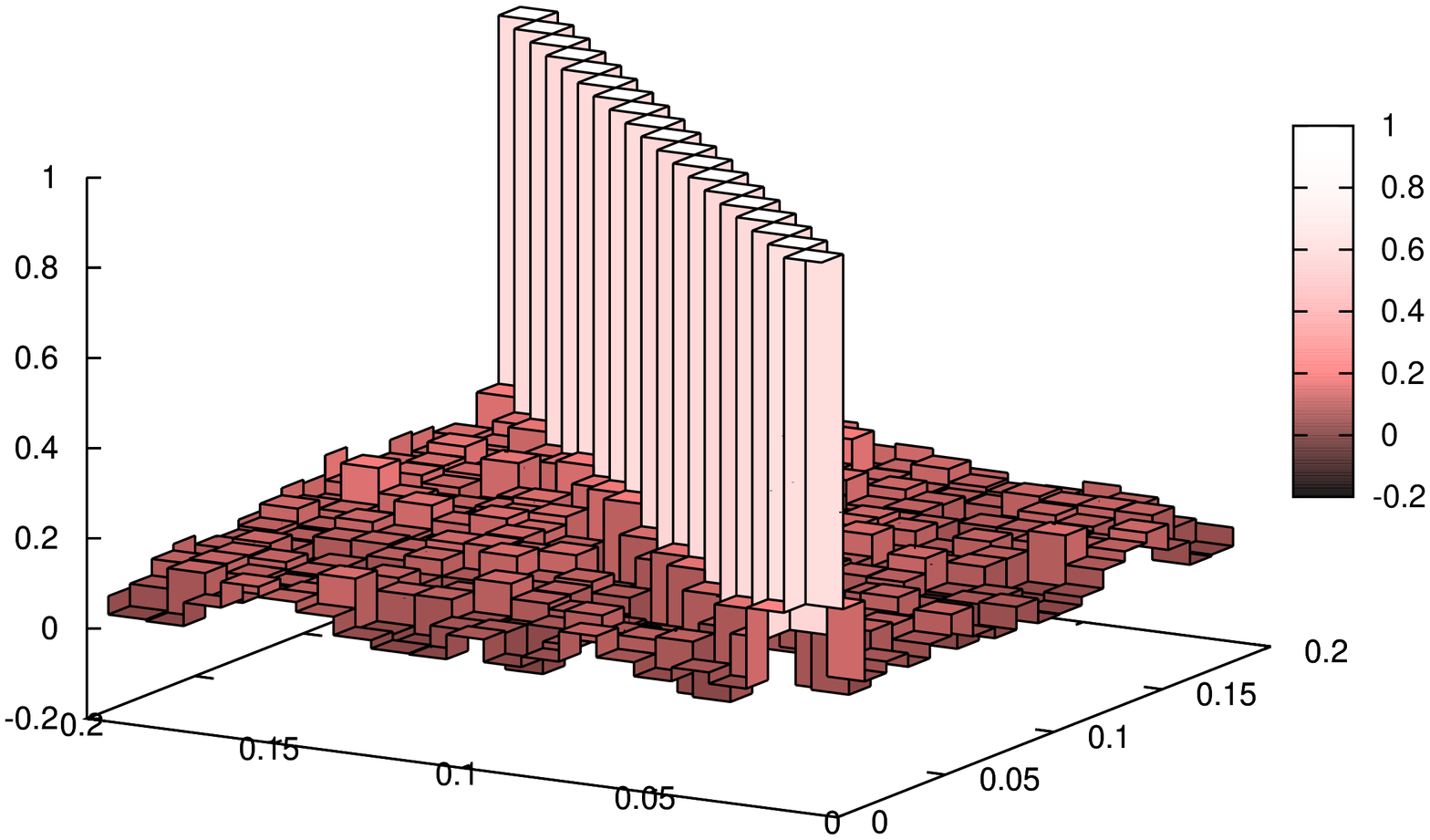}
&
    \includegraphics[width=0.22\textwidth,angle=270]{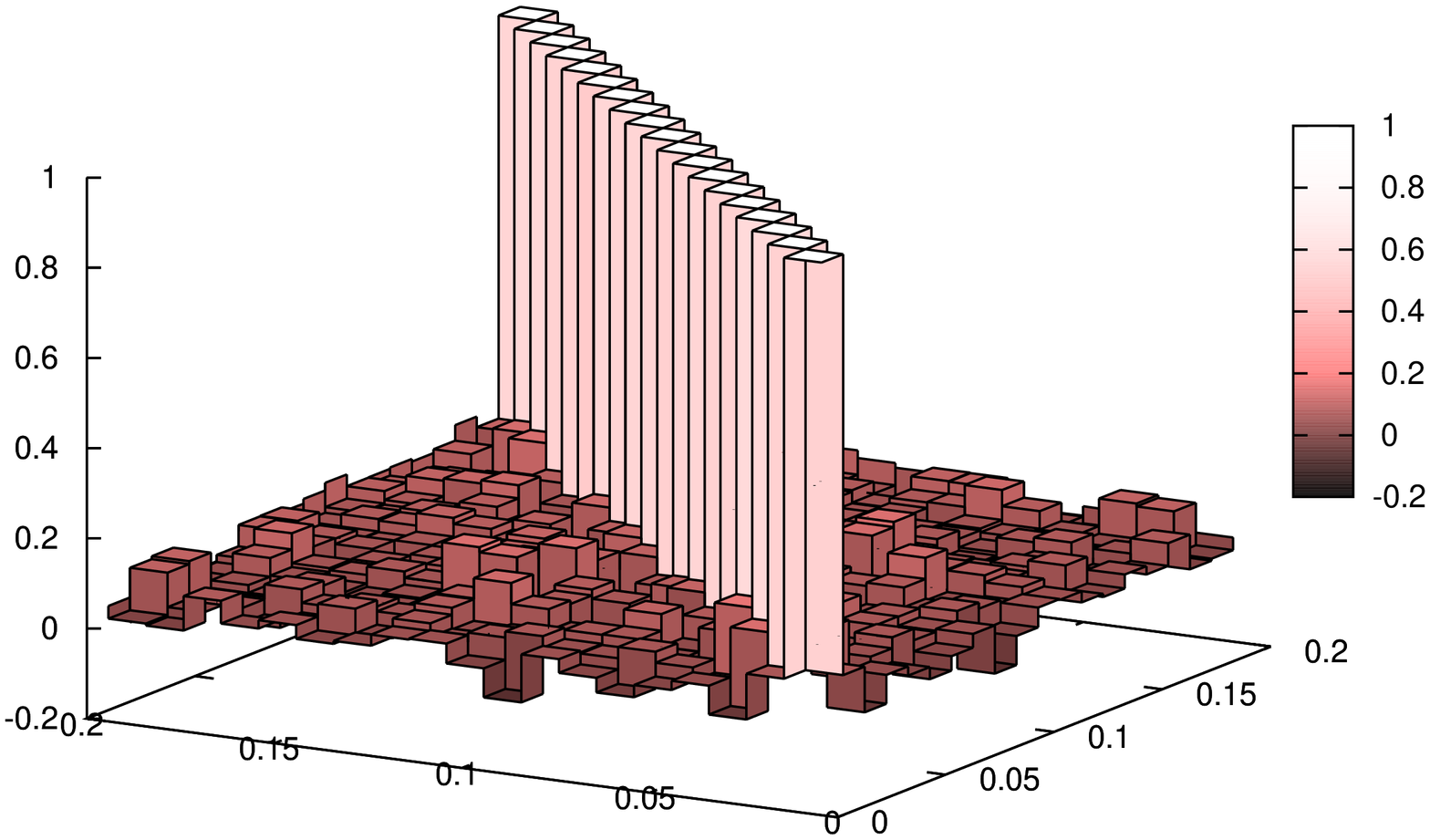}
&
    \includegraphics[width=0.22\textwidth,angle=270]{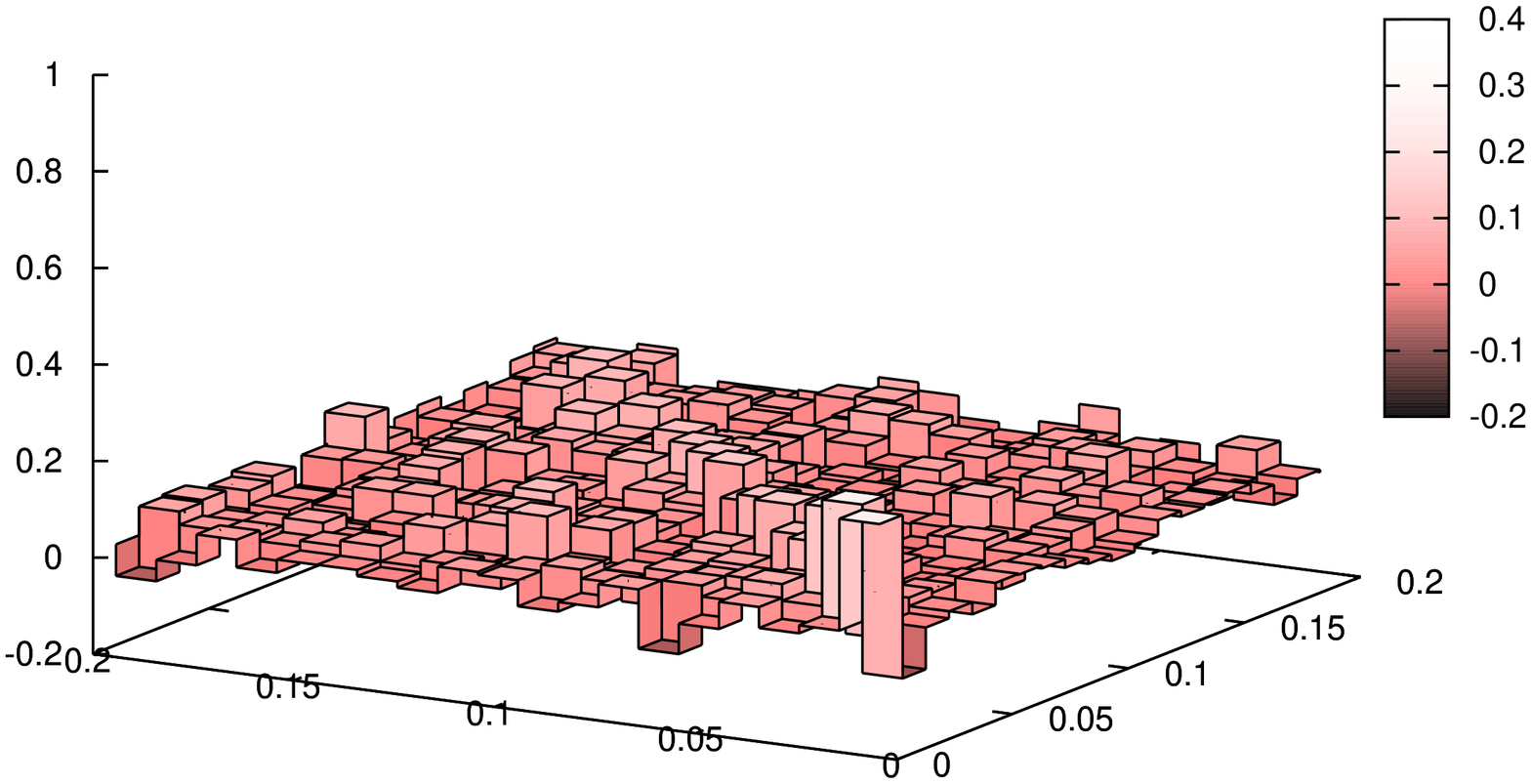}
\end{tabular}
\end{center}
\caption{Same as figure \ref{fig:corrP0} but with the {\it deconvolved}
power spectra.
\label{fig:corrP0dec}
}
\end{figure}

Figure \ref{fig:corrP0} shows the correlation matrices
of the {\it convolved} power spectra for
$\ell=\ell'=0$ (left panel), $\ell=\ell'=2$ (center panel) and
$\ell=0,\ell'=2$ (right panel), respectively, from the mock catalogue 
(A). Here, the full sample is divided into 18 subsamples, 
whose mean survey area is $397$ square degrees.
On the other hand, Figure \ref{fig:corrP0dec} shows the
correlation matrices of the {\it deconvolved} power spectra
for $\ell=\ell'=0$ (left panel), $\ell=\ell'=2$ (center panel)
and $\ell=0, \ell'=2$ (right panel)
respectively, from the mock catalogue (A).
The narrower subsample has a window function with a broader width, 
which makes the power spectrum at different wavenumbers more strongly 
correlated. 
One can see that the errors of the power spectra at different 
wavenumbers are correlated in figure \ref{fig:corrP0}, while practically 
de-correlated in figure \ref{fig:corrP0dec}. 
The off-diagonal components of figure \ref{fig:corrP0dec} are 
much smaller than those of figure \ref{fig:corrP0}.

\subsection{Baryon acoustic oscillation}
In this subsection, we discuss the effect of deconvolution
on the BAO signature. In general, the BAO signature in a convolved 
power spectrum is smoothed compared with the original power spectrum. 
Then, it is expected that the BAO signature of the deconvolved 
power spectrum has a larger amplitude compared to that of 
the convolved power 
spectrum. 

\begin{figure}[ht]
\begin{center}
\includegraphics[width=100mm, height=100mm]{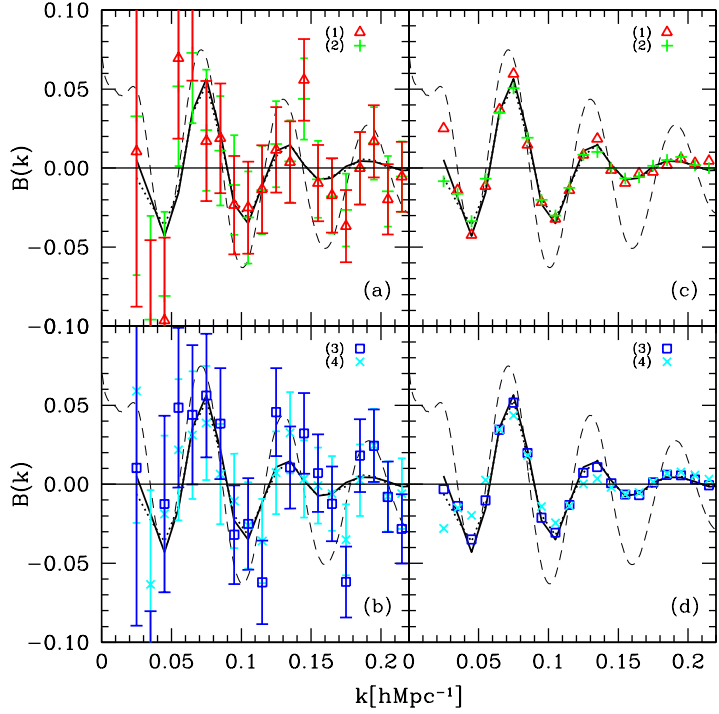}
\end{center}
\caption{
Baryon acoustic oscillation signature of the convolved and deconvolved
power spectra from the LRG sample (left panels) and $1000$ mock
catalogues (B) (right panels), respectively. 
In each panel, the solid curve and the dotted curve is from the
deconvolved spectrum and the convolved spectrum of the ideal
catalogues (C). 
(1) and (2) are obtained 
using the full sample without division into subsamples. 
(1) denoted by $\triangle$ is from the deconvolved power spectrum 
$P_0(k)$, while (2) denoted by $+$ is from the
convolved spectrum $P^{\rm conv}_0(k)$.
The results labelled by (3) and (4) are obtained using $18$
subsamples divided from the full sample. 
(3) denoted by $\square$ is from the deconvolved power spectrum, 
while (4) denoted by $\times$ is from the convolved spectrum, 
respectively.
The dashed curve is the theoretical curve in the linear theory. 
\label{fig:baosignature}
}
\end{figure}

Figure \ref{fig:baosignature} shows the BAO signature of the 
convolved and deconvolved power spectra from the LRG sample 
(left panels) and from $1000$ catalogues (B) (right panels), 
respectively.
In each panel, the solid curve and the dotted curve is the
BAO signature from the deconvolved spectrum and the 
convolved spectrum, respectively, obtained with 
$1000$ ideal catalogues (C). The dashed curve is 
the theoretical curve in the linear theory.

The results labelled by (1) and (2) are obtained
using the full sample without division into subsamples. 
(1) is from the deconvolved power spectrum $P_0(k)$, while (2) 
is the convolved spectrum $P^{\rm conv}_0(k)$. The BAO signature 
is obtained using the no-wiggle power spectrum multiplied by 
a function of the form $a+bk+ck^2+d/k^2$, where $a$, $b$, $c$, and
$d$ are fitting parameters (see ref.\citen{Nakamura} for details).
The errors in the left panels are large, but the following features 
can be seen in the right panels (the results from $1000$ mock 
catalogues (B)).
The difference between (1) and (2) is small, which means that the 
window effect is not very influential in this case because the 
survey  volume of the full SDSS LRG sample is large.
The difference between (1) and the solid curve and the dotted 
curve is also small. This means that the BAO signature of the 
deconvolved spectrum (1) well agrees with the original BAO 
signature.

The results labelled by (3) and (4) are obtained using $18$
subsamples divided from the full sample. 
(3) is from the deconvolved power spectrum, while (4)
from the convolved one, respectively.
The difference between (3) and (4) is not negligible at 
the peaks and the troughs due to the window effect, 
while the difference  between (3) and the solid curve and the
dotted curve is small. This means that 
the deconvolution is successful and recovers the BAO signature, 
although the window effect is substantial, and degrades the BAO 
signature when the survey volume is small. 
Practically, we need not divide a full sample into subsamples
to obtain the BAO signature from the monopole spectrum. 
Then, the window effect on the BAO analysis seems to be small 
compared with errors. However, in a future precise 
measurement of the BAO signature, the prescription for 
the window effect might become important.

\section{Summary and conclusions}
Large galaxy surveys are promising tools for the future dark energy 
research.
In a redshift survey, the power spectrum analysis is very fundamental 
for extracting the BAO signature. Similarly,
tests of the general relativity on the scales of cosmology are becoming
important related to this topic in cosmology. 
This is because a modification of the gravity theory is an alternative way to 
explain the cosmic acceleration, instead of introducing the dark energy 
component. The key to distinguish between the dark energy and the 
modified gravity is the evolution of the cosmological perturbations.
The redshift-space distortion caused by the peculiar motions of the 
galaxies, provides us a useful chance to 
test gravity on the scale of cosmology \cite{Yamamoto08,Yamamoto10}. 
Here the multipole power spectrum plays a key role. 

In this paper, we developed a new method to measure the deconvolved 
power spectra.  By applying it to the SDSS DR7 LRG sample as well as to
the mock catalogues, we demonstrated that the scheme works well. 
By the deconvolution, 
we can get the power spectra whose amplitudes are properly recovered.  
This will be essential in measuring the redshift-space distortions.
The covariance matrices of the deconvolved power spectra are 
practically diagonal. This is also useful in the study of the 
BAO signature. 
Our method matches with the fast Fourier transform, which saves 
computation time significantly. 

As mentioned at the end of the subsection 3.1, the limitation 
of the distant observer approximation might cause a small 
deviation of the quadrupole power spectrum when compared 
with theoretical prediction derived on the basis of 
the distant observer approximation. For a precise treatment, 
the framework of the spherical redshift-space distortion is 
necessary \cite{SD,SD2}.  However, this problem might be 
softer for a galaxy sample at higher redshift, because 
a volume within a small angular area can be large.

\vspace{0.3cm}
{\it Acknowledgement}
This work was supported by Japan Society for Promotion
of Science (JSPS) Grants-in-Aid
for Scientific Research (Nos.~21540270,~21244033).
This work is also supported by JSPS 
Core-to-Core Program ``International Research 
Network for Dark Energy''.
We thank G. Nakamura and A. Taruya for useful discussions.


\begin{thebibliography}{999} 
\bibitem{Dodelson} S. Dodelson, {\it Modern Cosmology} (Academic Press, 2003)
\bibitem{Tegmark06} M. Tegmark, et~al., Phys. Rev. D {\bf 74} (2006), 123507 
\bibitem{Hutsia}
 G. H\"{u}tsi, Astron. Astrophys. {\bf 449} (2006), 891 
\bibitem{Hutsib}
 G. H\"{u}tsi, Astron. Astrophys. {\bf 459} (2006), 375 
\bibitem{Percival} 
W. Percival, et al., Astrophys. J. {\bf 657} (2007), 645; 
W. Percival, et al., Mon. Not. Roy. Astron. Soc. {\bf 381} (2007), 1053; 
W. Percival, et al., Mon. Not. Roy. Astron. Soc. {\bf 401} (2010), 2148 
\bibitem{Linder} 
E. V. Linder, Astropart. Phys. {\bf 29} (2008), 336 
\bibitem{Guzzo}
 L. Guzzo et~al., Nature {\bf 451} (2008), 541 
\bibitem{Yamamoto08}
 K. Yamamoto, T. Sato and G. H\"{u}tsi Prog. Theor. Phys.~{\bf 120} (2008), 
609 
\bibitem{FKP94} 
 H. A. Feldman, N. Kaiser and J. A. Peacock,  
Astrophys. J. {\bf 426} (1994), 23 
\bibitem{SatoW2} T. Sato, G. H\"utsi, G. Nakamura and K. Yamamoto,
(unpublished)
\bibitem{Fisher93} K. Fisher, M. Davis, M. A. Strauss, A. Yahil 
and J. P. Muchra, Astrophys. J. {\bf 402} (1993), 42 
\bibitem{CFW} S. Cole, K. Fisher and D. H. Weinberg,  Mon. Not. Roy. 
Astron. Soc. {\bf 267} (1994), 785 
\bibitem{Hamilton} A. J. S. Hamilton, in {\it The Evolving Universe} 
(Dordrecht: Kluwer Academic Publishers, 1998), 185
\bibitem{YNKBN} K. Yamamoto, M. Nakamichi, A. Kamino, B. A. Bassett and
H. Nishioka, Publ. Astron. Soc. Japan {\bf 58} (2006), 93
\bibitem{OutramI}
 P. J. Outram, et al., Mon. Not. Roy. Astron. Soc. {\bf 328} (2001), 174 
\bibitem{OutramII}
 P. J. Outram, et al., Mon. Not. Roy. Astron. Soc. {\bf 348} (2004), 745 
\bibitem{Yamamoto10} K. Yamamoto, G. Nakamura, G. H\"utsi, T. Narikawa 
and T. Sato, Phys. Rev. {\bf D 81} (2010), 103517 
\bibitem{Nakamura}
 G. Nakamura, G. H\"{u}tsi, T. Sato and K. Yamamoto, Phys. Rev. D {\bf 80} (2009), 123524 
\bibitem{SD}
 A. J. S. Hamilton and M. Culhane, 
 Mon. Not. Roy. Astron. Soc. {\bf 278} (1996), 73
\bibitem{SD2}
 A.  Raccanelli, L. Samushia and W. J. Percival,  
 Mon. Not. Roy. Astron. Soc., in press, arXiv:1006.1652 
\end{thebibliography}
\end{document}